# Beyond the single-file fluid limit using transfer matrix method: Exact results for confined parallel hard squares

Péter Gurin and Szabolcs Varga

*Institute of Physics and Mechatronics, University of Pannonia, PO Box 158, Veszprém, H-8201 Hungary*

Number of pages: 27 (including figures)

Figures: 5

**Abstract**

We extend the transfer matrix method of one-dimensional hard core fluids placed between confining walls for that case where the particles can pass each other and at most two layers can form. We derive an eigenvalue equation for a quasi-one-dimensional system of hard squares confined between two parallel walls, where the pore width is between $\sigma$ and $3\sigma$ ($\sigma$ is the side length of the square). The exact equation of state and the nearest neighbor distribution functions show three different structures: a fluid phase with one layer, a fluid phase with two layers and a solid-like structure where the fluid layers are strongly correlated. The structural transition between differently ordered fluids develops continuously with increasing density, i.e. no thermodynamic phase transition occurs. The high density structure of the system consists of clusters with two layers which are broken with particles staying in the middle of the pore.

# 1 Introduction

Phase behavior of colloidal systems can be altered substantially in geometrically restricted environments [1-3]. Well-known examples are the crossover from normal diffusion to single-file diffusion [4-7], the replacement of true phase transitions by the continuous structural changes [8-11] and the confinement induced inhomogeneous density distributions, layering and depletion attractions [12]. Further curiosities are the wall induced helical arrangement of spherical particles in cylindrical pore [13-15] and the density anomaly taking place with cooling down in cylindrically confined fluids [16].

Theoretical study of quasi-one-dimensional fluids is still fascinating topic. The reason for this is that we can gain information about freezing and glass transition properties of two- and three-dimensional fluids with less computational efforts [17]. In addition to this, the effect of cylindrical and slit-like confinements can be studied on structural and dynamical properties of confined colloidal systems. Two types of models are common in the literature: the particles are allowed to move freely along a longitudinal direction in both models (the system is bulk in one direction), but either a periodic boundary condition or a geometrical confinement is used in the transversal direction. The first model is devised to study the bulk properties, while the second one is for understanding the effect of confinement. The spherical and the polyhedral geometrical shapes are the most common representations of the particles in the theoretical studies due the sudden development of colloidal science in synthesis of colloids with these shapes [18-21].

Theoretical development in the description of quasi-one-dimensional fluids dates back to the exact equation of states of one-dimensional hards rods due to Tonks [22]. Later, an eigenvalue equation was derived for quasi-one-dimensional systems where not only nearest neighbor interactions are present [23]. Percus and Zhang developed a similar formalism for quasi-one-dimensional hard square fluid, where the particles can pass each other, i.e. first and second neighbor interactions are present [24]. Kofke and Post presented an alternative way to obtain Barker's eigenvalue equation, where hard particles interact only with their nearest neighbors [25]. This method was the so-called transfer matrix method of continuum systems, which was devised earlier for one-dimensional fluid of anisotropic particles with continuous positional and rotational freedom (1D rotor) [26]. The density dependence of the longitudinal and transversal pressure was possible to examine in several systems such as the confined hard spheres and hard disks due to the transfer matrix method and virial expansion [27-31]. Recently the transfer matrix method has been proved successful to determine the positional correlation length [32], pair correlation function [33] and structure factor [34]. It also serves information about the dynamics, possible glassy states and jamming behavior of confined hard disks [34]. In a series of papers, Bowles et al. have managed to prove that a fragile-strong fluid structural rearrangement takes place in the system of confined hard disks, which is located at the maximum of isobaric heat capacity [35-38]. It has been shown very recently by Godfrey and Moore [39] that the positional correlation length is practically the spacing between the neighboring defects, which are responsible for the presence of jammed states. They have also managed to extend the transfer matrix method for slightly wider pores and examine the glassy behavior of hard disks, where both first and second neighbor interactions are present, but the disks cannot pass each other [40].

In this work we extend the transfer matrix formalism of nearest neighbor interacting single-file hard body fluids [25] for those cases when second neighbor interactions are present and the single-file condition violates. These conditions are accompanied by the fact that the particles can pass each other in the confined geometry. We apply the theory for parallel hard squares which are confined between two parallel hard walls. We show that the violation of the single-file condition dramatically alters the phase behavior of the hard squares. For example, the equation of state of one-dimensional hard rods (Tonks equation) deviates substantially from the equation of state coming from the transfer matrix method. Furthermore instead of only one layer, two layers can be formed with hard squares between confining hard walls, where the strong correlation between the layers is an indicator of the crystallization into rectangular lattice.

The paper is organized as follows. In Sec. II the transfer matrix method is presented for quasi-one-dimensional confined hard bodies. An eigenvalue equation is derived for general particle shape and solved for parallel hard squares in very narrow pores. Numerical details of the solution of the eigenvalue equation are also discussed for wider pores in this section. Results for the equation of state and positional distribution functions are presented and discussed at two pore widths in Sec. III. The paper concludes with some remarks in Sec. IV.

## 2  Transfer matrix method for first and second neighbor interacting systems

We consider the system of two-dimensional hard core particles confined between two parallel walls. The geometrical confinement allows the particles to pass each other, but the particles can interact with only first and second neighbors, which will be discussed later. The pair interaction between two particles $i$ and $j$ is purely hard, i.e.

$$u_{ij} = \begin{cases} \infty & \text{for overlap} \\ 0 & \text{otherwise} \end{cases}, \qquad (1)$$

and the hard repulsive wall-particle interaction restricts the transverse position of the particles to be in a well-defined interval. We use an effective pore width ($W$) which restricts the transverse position of the particle's centre to be between $-W/2$ and $W/2$, but the particles can be anywhere along the longitudinal direction. Note that $W$ cannot be too large to fulfill the first and second neighbor interaction condition. At this stage we develop a general theory by not specifying the shape of the particles. This means that our theory can be applied for multi-faceted and circular shapes, but the accessible range of *W* is changing from shape to shape and the condition for mutual passage cannot be held. For example the violation of single-file condition in confined hard disks is accompanied by the presence of third neighbor interactions.

The most convenient statistical ensemble for continuum confined hard body systems is the isobaric one, where we fix the longitudinal one-dimensional pressure (*P*) and the pore width (*W*), while the length of the system (*L*) is allowed to fluctuate along the longitudinal direction. In this ensemble the configurational part of the isobaric partition function of any system can be written as [41]

$$Z_{NPT} = \frac{1}{N!} \int dL \prod_{i=1}^{N} \int d\vec{r}_i e^{-\beta u_{tot} - \beta PL} \quad , \tag{2}$$

where *N* is the number of particles, $\beta = \frac{1}{k_B T}$ , $\vec{r}_i$ is abbreviation for the tranverse and longitudinal positions of particle *i*, *L* is the longitudinal length of the system and $u_{tot}$ is the total pair potential (sum of the pair potentials). To evaluate the isobaric partition function we introduce new variables and group the neighboring particles together such that *N*/2 pairs (dimers) constitute the whole system. One pair consists of two longitudinally neighboring particles, which is not same with a pair with shortest distance. Fig. 1 shows the pairing of the particles. One can see that $x_i$ denotes the longitudinal distance between the particles of dimer *i*, while $y_{i,1}$ and $y_{i,2}$ are the transverse positions of the particle 1 and 2 of the dimer *i*. We also use $X_{i,i+1}$, which is the longitudinal distance between the centres of neighboring dimers *i* and *i*+1. In the followings we use the periodic boundary condition in longitudinal direction in a sense that the particle labelled by *N*+1 is identical with the first particle. Keeping in mind that only first and second neighbor interactions are taking place, i.e. $u_{tot} = \sum_{i=1}^{N}(u_{i,i+1} + u_{i,i+2})$, apart from an unimportant factor the partition function can be rewritten into the following form:

$$Z_{NPT} = \prod_{i=1}^{N/2} \left( \int_{-W/2}^{W/2} dy_{i,1} \int_{-W/2}^{W/2} dy_{i,2} \int_{\sigma(y_{i,1},y_{i,2})}^{\infty} dx_i \int_{\sigma_{i,i+1}^*}^{\infty} dX_{i,i+1} \right) e^{-\beta P \sum_{j=1}^{N/2} X_{j,j+1}}, \quad (3)$$

where $\sigma(y_{i,1}, y_{i,2})$ is the longitudinal contact distance between two neighbors if the neighboring particles are in $y_{i,1}$ and $y_{i,2}$ transverse positions. The contact distance between the centres of neighboring dimers appearing in Eq. (3) is given by

$$\sigma_{i,i+1}^* = \max \begin{Bmatrix} \sigma(y_{i,1}, y_{i+1,1}) - x_i/2 + x_{i+1}/2 \\ \sigma(y_{i,2}, y_{i+1,2}) + x_i/2 - x_{i+1}/2 \\ \sigma(y_{i,2}, y_{i+1,1}) + x_i/2 + x_{i+1}/2 \end{Bmatrix}. \quad (4)$$

In this expression the first two rows give the contact distance between the pairs if the second neighbors are in contact, while the third one is for the first neighbor contact. Note that $N!$ term is missing in Eq. (3), because we integrate only over that region of the configurational space, where the longitudinal coordinates of the particles are kept in a given order instead of integrating over all possible values of the positions. The validity of this procedure is due to the permutation symmetry of the system because all configurations of the system can be obtained from our used order by a permutation. At this point it is worth mentioning that the number of integrations in Eq. (3) is less by one than in Eq. (2). However it can be proved that the missing integration does not have effect on the results in the thermodynamic limit. We can perform further simplification in Eq. (3), because there is no coupling between contact distances ($X_{i,i+1}$). Integrating out $X_{i,i+1}$ dependence in Eq. (3) we get that

$$Z_{NPT} = \int_{-W/2}^{W/2} dy_{1,1} \int_{-W/2}^{W/2} dy_{1,2} ... \int_{-W/2}^{W/2} dy_{N/2,1} \int_{-W/2}^{W/2} dy_{N/2,2} \int_{\sigma(y_{1,1},y_{1,2})}^{\infty} dx_1 ... \int_{\sigma(y_{N/2,1},y_{N/2,2})}^{\infty} dx_{N/2} \frac{e^{-\beta P \sigma_{1,2}^*}}{\beta P} ...... \frac{e^{-\beta P \sigma_{N/2,1}^*}}{\beta P} \quad (5)$$

This expression can be rewritten into a more compact form: $Z_{NPT} = TrK^{N/2}$, where $Tr$ is the trace of the integral operator $K^{N/2}$ and $K$ is defined by the kernel $K(y_{1,1}, y_{1,2}, x_1; y_{2,1}, y_{2,2}, x_2) = \exp(-\beta P \sigma^*)/\beta P$. In the thermodynamic limit ($N \to \infty$) one can show that $Z_{NPT} = \lambda_0^{N/2}$, where $\lambda_0$ is the largest eigenvalue of $K$. In detail $\lambda_0$ is the largest solution of the following eigenvalue equation

$$\int_{-W/2}^{W/2} dy_{2,1} \int_{-W/2}^{W/2} dy_{2,2} \int_{\sigma(y_{2,1}, y_{2,2})}^{\infty} dx_2 K(y_{1,1}, y_{1,2}, x_1; y_{2,1}, y_{2,2}, x_2) \psi_k(y_{2,1}, y_{2,2}, x_2) = \lambda_k \psi_k(y_{1,1}, y_{1,2}, x_1), \quad (6)$$

where $\psi_k$ is the eigenfunction of the operator for a pair and $k$ is integer ($k=0,1,2,3...$). The physically meaningful solution of Eq. (6) is the largest eigenvalue ($\lambda_0$) and the corresponding eigenfunction ($\psi_0$). Using the largest eigenvalue we can determine the Gibbs free energy as follows $\beta G/N = -0.5 \ln \lambda_0$, while a positional probability distribution function of a nearest neighbor pair can be constructed from the eigenfunction as follows $f = \psi_0(y, y', x) \psi_0(y', y, x)$ if $\psi_0(y, y', x)$ satisfies the following normalization condition:

$$\int_{-W/2}^{W/2} dy \int_{-W/2}^{W/2} dy' \int_{\sigma(y,y')}^{\infty} dx \psi_0(y, y', x) \psi_0(y', y, x) = 1, \quad (7)$$

where $y$ and $y'$ are the transversal positions of the neighboring particles and $x$ is the distance between them. Note that $\sigma(y, y')$ longitudinal contact distance ensures that the particles are not allowed to overlap with each other. The equation of state can obtained directly from the Gibbs free energy as follows $1/\rho = \partial(\beta G/N)/\partial(\beta P)$, where $\rho = N/<L>$ is the linear density. The other route for the equation of state is to determine the average distance between neighboring particles from $\int_{-W/2}^{W/2} dy \int_{-W/2}^{W/2} dy' \int_{\sigma(y,y')}^{\infty} dx \psi_0(y, y', x) x \psi_0(y', y, x) = \langle x \rangle$ and use the fact that there is only one particle on the average distance, i.e. $\rho = N/<L> = 1/\langle x \rangle$.

At this point it is worth showing that Eq. (6) reproduces the well-known Tonks equation of the one-dimensional hard rods [22]. Restricting the particles into such narrow pores that the longitudinal contact distance is constant ($\sigma$) Eq. (6) can be written as

$$\int_{-W/2}^{W/2} dy_{21} \int_{-W/2}^{W/2} dy_{22} \int_{\sigma}^{\infty} dx_2 \frac{e^{-\beta P \sigma_{1,2}^*}}{\beta P} \psi_k(y_{2,1}, y_{2,2}, x_2) = \lambda_k \psi_k(y_{1,1}, y_{1,2}, x_1). \quad (8)$$

where $\sigma_{1,2}^* = \sigma + x_1/2 + x_2/2$ because only nearest neighbor interactions take place (see Eq. (4)). Note that there is no $y$ dependence in the kernel, i.e. the eigenfunction cannot depend on $y$. It is easy to show that the solution of the above equation for the eigenfunction and eigenvalue are $\psi_0(x) = \theta(x-\sigma)\alpha e^{-\beta Px/2}$ and $\lambda_0 = \left(W e^{-\beta P \sigma}/\beta P\right)^2$, where $\alpha$ is a normalization constant and $\theta$ is the Heaviside step function ($\theta(x)=1$ for $x \geq 0$ and $\theta(x)=0$ for $x<0$). Both routes of the equation of state, i.e. $1/\rho = \partial(\beta G/N)/\partial \beta P$ and $\rho = 1/\langle x \rangle$ equations give the same results for the linear density, namely $1/\rho = \sigma + 1/\beta P$.

Now we consider the system of parallel hard squares with side length $\sigma$ in a narrow pore, where only first and second neighbor interactions are allowed (see Fig. 2). This criterion is satisfied if the pore width is not larger than $3\sigma$, i.e. $W<2\sigma$. One can realize that the longitudinal contact distance $\sigma(y,y')$ has only two values, it is zero if $|y-y'| \geq \sigma$ and it is $\sigma$ otherwise. Exploiting this fact in Eq. (4), Eq. (6) can be rewritten as

$$\lambda_0 \overline{\psi}_0(y_1, x_1) = \frac{e^{-\beta P \sigma}}{\beta P} \int_{-W/2}^{W/2} dy_2 \left[ A(y_1, y_2) \left( e^{-\beta P x_1/2} \int_0^{x_1} dx_2 e^{\beta P x_2/2} \overline{\psi}_0(y_2, x_2) + e^{\beta P x_1/2} \int_{x_1}^{\sigma} dx_2 e^{-\beta P x_2/2} \overline{\psi}_0(y_2, x_2) \right) \right.$$
$$\left. + (a(y_2) - A(y_1, y_2)) e^{-\beta P x_1/2} \int_0^{\sigma} dx_2 e^{-\beta P x_2/2} \overline{\psi}_0(y_2, x_2) \right]$$
$$+ \frac{e^{-1.5\beta P \sigma}}{(\beta P)^2} \left[ e^{\beta P x_1/2} a(y_1) + e^{-\beta P x_1/2}(W - a(y_1)) \right] \int_{-W/2}^{W/2} dy \overline{\psi}_0(y, \sigma),$$

(9)

where $a(y)$ and $A(y, y')$ are defined as

$$a(y) = \theta(|y| + W/2 - \sigma)(|y| + W/2 - \sigma), \tag{10}$$

and

$$A(y, y') = [\min(|y|, |y'|) + W/2 - \sigma][\theta(\min(y, y') + W/2 - \sigma) + \theta(-\max(y, y') + W/2 - \sigma)]. \tag{11}$$

Both $a$ and $A$ have simple geometrical meaning, because $a(y)$ is simple a free length not excluded by a particle staying at $y$ position for the other particle staying in the same longitudinal position, while $A(y, y')$ is also a free length not excluded by neither a particle staying at $y$ nor by another particle staying at $y'$. The connection between $\psi$ and $\bar{\psi}$ is given by

$$\psi_0(y, y', x) = \begin{cases} \theta(|y - y'| - \sigma)\bar{\psi}_0(y', x) & \text{if } x < \sigma \\ \bar{\psi}_0(y', \sigma)e^{-\beta P(x - \sigma)/2} & \text{if } x \geq \sigma \end{cases}. \tag{12}$$

Note that the Tonks-like exponential decay of the eigenfunction for $x > \sigma$ survives for wider pores, too. We have solved Eq. (9) at a given pore width ($W$) and longitudinal pressure ($P$) by iteration using the normalization condition of the eigenfunction (Eq. (7)). The integrations are performed numerically using the trapezoidal quadrature for both longitudinal and transverse variables. For wider pores ($W > 1.5\ \sigma$) $\Delta x/\sigma = 0.01$ and $\Delta y/\sigma = 0.01$ grid sizes for $x$ and $y$ variables have proved sufficient even in the vicinity of close packing density. However, smaller grid sizes must be used for narrow pores ($W < 1.5\ \sigma$) at high pressure because of sudden change of the eigenfunction in $-W/2 < y < W/2 - \sigma$ and $-W/2 + \sigma < y < W/2$ intervals. We overcome this problem by using finer grid size ($\Delta y/\sigma = 0.001$) in the above two intervals. Using the above grid sizes the energy route $(1/\rho = \partial(\beta G/N)/\partial \beta P)$ and the average distance route $(\rho = 1/\langle x \rangle)$ give the same density at a given longitudinal pressure. However, we must mention that the energy route is less sensitive to applied grid size than the average distance one. Finally note that Eq.(9) is exact only for $W < 2\sigma$, because the possible effect of third neighbor interactions is not included into our transfer matrix formalism, which can be crucial at the description of close packing structures of $W > 2\sigma$ cases.

## 3  Results and discussion

The hard squares form a single-file and only nearest neighbor interactions occur for $W < \sigma$. In this case the thermodynamic and structural properties of hard squares are identical with those of Tonks gas of one-dimensional hard rods [22]. In the case of $\sigma < W < 2\sigma$ one or two fluid layers can accommodate between the confining walls where in addition to the first neighbor interactions second neighbor ones are also present. We go beyond the single-file fluid and Tonks-gas limits by presenting our exact numerical results for $W = 1.08\sigma$ and $W = 1.92\sigma$ pore widths. The first one is just wider than the limiting single-file fluid pore width, while the second one is very close to the upper limit of the pore width $(W_{max} = 2\sigma)$ where third neighbor interactions start to be present.

We have observed two different structures: one is fluid-like, while the other is solid-like. In the fluid-like structure only short ranged positional order takes place, where the particles like to stay in the vicinity of the wall to maximize the available free area for the rest of the particles (upper panel of Fig. 2). In the case of solid-like order two fluid layers form, one is located at the lower wall, while the other is at the upper one. These two fluid layers become longitudinally correlated with increasing pressure in such a way that rectangular lattice structure develops (lower panel of Fig. 2.). Phase transitions do not occur in our systems.

To understand the continuous structural change happening with the increasing density in our system we determine the equation of state, the longitudinal and transversal positional distribution functions and the mole fractions of the squares disrupting the two-fluid-layer structure (defect's mole fraction). The longitudinal distribution function of a square around its nearest neighbor can be determined from the eigenfunction as follows

$$<f_x(x)> = \int_{-W/2}^{W/2} dy \int_{-W/2}^{W/2} dy' \psi_0(y,y',x) \, \psi_0(y',y,x) \,. \qquad (13)$$

The wall-to-wall transversal distribution function of the squares can be obtained similarly from

$$<f_y(y)> = \int_{-W/2}^{W/2} dy' \int_0^\infty dx \, \psi_0(y,y',x) \, \psi_0(y',y,x). \qquad (14)$$

To examine how the structure transforms from fluid-like order to solid one we measure the fraction of the defect particles (squares) staying between $W/2-\sigma$ and $-W/2+\sigma$ transversal positions, which corresponds to those cases when the formation of two fluid layers is forbidden. This mole fraction is calculated from

$$\langle X_1 \rangle = \int_{W/2-\sigma}^{-W/2+\sigma} \langle f_y(y) \rangle dy \,. \qquad (15)$$

Other stringent indicator of the formation of two-fluid-layer structure is the fraction of those dimers where both particles disrupt the possible two-fluid structure. This dimer (defect pair) mole fraction is given by

$$<X_2> = \int_{W/2-\sigma}^{-W/2+\sigma} dy \int_{W/2-\sigma}^{-W/2+\sigma} dy' \int_0^\infty dx \, \psi_0(y,y',x) \, \psi_0(y',y,x). \qquad (16)$$

The equation of state is presented in Figs. 3a and 3b for $W = 1.08\sigma$ and $W = 1.92\sigma$, respectively, where both the pressure and the density are two-dimensional: $P^* = \beta P \sigma^2/(\sigma+W)$ is the dimensionless pressure, while $\eta = N\sigma^2/A$ is the packing fraction and $A = (\sigma+W)\langle L \rangle$ is the area of the system. In the narrower pore $(W = 1.08\sigma)$ we can distinguish three different parts in the curve (Fig. 3a): 1) a low density fluid phase with one fluid layer, 2) a middle density fluid phase with one and two fluid layers and 3) a high density fluid phase with strong spatial correlations between the two fluid layers. In the low density region where $\eta < 0.3$, the results of the transfer matrix method (solid curve) and those of Tonks-equation for hard rods (dashed curve) are practically identical. This is not surprising because the Tonks-equation, which is given by $\beta P = \rho/(1-\rho\sigma)$, is exact for $W < \sigma$. In this region the fraction of squares staying in the centre of pore is very large (see Fig 4) and almost equal to the limiting ideal-gas value given by $\langle X_1 \rangle_{ideal} = (2\sigma - W)/W$. At $\eta = 0.3$ one can see that $\langle X_1 \rangle \approx 0.81$, while $\langle X_1 \rangle_{ideal} \approx 0.85$. The fraction of two neighboring squares staying in the central part of the pore is also very high, for example $\langle X_2 \rangle \approx 0.68$ at $\eta = 0.3$ which is close to the ideal-limit value given by $\langle X_2 \rangle_{ideal} = \langle X_1 \rangle_{ideal}^2 \approx 0.73$. This shows that there are lots of neighbors in the centre of the pore and they form one-dimensional Tonks clusters with varying lengths along the pore. This Tonks-fluid behavior of the system is not affected by the presence of the particles located at the vicinity of walls at $\eta = 0.3$, where the mole fraction of squares is about 0.19. In this low-density region the effect of second neighbor interactions is obviously negligible. In the middle density region $(0.3 < \eta < 0.8)$ the decrease in $dP/d\eta$ indicates that structural changes must happen in the pore, i.e. the system start to behave differently from the Tonks-fluid. This manifests in the increasing difference between the pressure of Tonks-fluid and that of confined hard squares with increasing density. The highest value of the packing fraction at which it is possible to accommodate the pore with only one fluid layer is about 0.48. This means that the particles are forced to move into the directions of walls, two fluid layers start to develop at the walls and the centre of the pore becomes less and less occupied by increasing density. In this region both $\langle X_1 \rangle$ and $\langle X_2 \rangle$ decreases rapidly with increasing density, i.e. the system loses its one fluid nature and clusters with two layers become more and more dominant. At $\eta = 0.8$ $\langle X_2 \rangle$ is practically zero, i.e. it is hard to find two neighboring particles at the centre of the pore. Therefore only clusters of two fluid layers survive for $\eta > 0.8$, which are interrupted by few defect particles staying in the middle of the pore as $\langle X_1 \rangle$ is still not negligible. One can imagine that two Tonks fluids develop, one is at the upper wall, while the other at the lower wall. These two fluids would be independent, i.e. their structures are not correlated and crystallization does not take place with increasing density. To check this scenario, the longitudinal nearest neighbor and the transversal positional distribution functions are plotted in Fig. 5. $\langle f_y \rangle$ shows that central part of pore becomes empty with increasing density, while $\langle f_x \rangle$ indicates that upper and the lower fluid layers are correlated. Two favorable positions can be seen from the plot of $\langle f_x \rangle$ for the neighboring particle around a given one: 1) it stays in the same longitudinal position (x=0), i.e. the two particles form a column or 2) it is located at $x=\sigma$ distance. This shows that clusters of rectangular lattices develop with increasing density and the clusters are interrupted by the defect particles staying the centre of the pore. The concentration of the defect particles decreases with density, i.e. the close packing structure of the system is simple rectangular lattice. Therefore the stabilization of the rectangular lattice can be attributed to the entropic gain taking place in vertical fluctuations, while the role of defect particles is to suppress the longitudinal fluctuations. The regular zigzag arrangement cannot be stabilized in this system, because the longitudinal fluctuations would be suppressed in higher extent.

Hard squares behave slightly differently in the wider pore ($W = 1.92\sigma$) as can be seen from the equation of state (Fig. 3b). No change can be seen in the shape of the curve and there is no sign of the formation of single fluid layer. The two equations of state (Tonks and the present one) meet practically only in the ideal gas limit. This can be attributed to the fact that the particles can pass each other easily and clusters with one layer cannot form along the pore. In practice the occupation of the pore with two fluid layers starts at very low density and no structural change occurs with increasing density. This is confirmed by the inset of Fig. 4, where one can see that the fraction of squares (dimers) in the central part of the pore is very low even in the ideal gas limit. One can gain some insight into the high-density structure by looking at the horizontal and transversal distribution functions shown in Fig. 5b. The smoothly varying $\langle f_y \rangle$ indicates that the particles can move easily in transverse direction even at the close packing limit. In addition to this it shows that one fluid layer is located at the upper wall, while the other one is at the lower wall. Practically there are no squares in the centre of the pore (see the case of $P^* = 5$). The correlation between the two fluid layers is examined by the nearest neighbor longitudinal distribution function (inset of Fig. 5b). One can see that the two layers are strongly correlated at high densities due to the large positional fluctuations along the transverse direction. The presence of very few defect particles in the centre does not effect substantially the correlation between the layers. Therefore the close packing structure of the wide pore is also rectangular lattice.

## 4 Conclusions

We have extended the transfer matrix method of quasi-one-dimensional fluids for wider pores where the particles can pass each other and at most two layers can form. We have derived an eigenvalue equation for the nearest neighbors, where the eigenfunction carries the information about the positional distribution of the nearest neighbors and the eigenfunction gives the Gibbs free energy of the system. It is shown that the pair distribution function is proportional to $\exp(-\beta P x)$ for $x \geq \sigma$ distances, where $\sigma$ is the largest length of the hard body. This makes possible to determine the longitudinal pressure of confined systems using MC simulation methods in canonical (NVT) ensemble by measuring the pair distribution function. The method can be used for several hard body shapes such as squares, rectangles and disks if only first and second neighbor interactions are present.

As a demonstration of the method we have examined the system of parallel hard squares confined between two parallel walls with $W = 1.08\sigma$ and $W = 1.92\sigma$ pore widths. Our results show that two fluid layers develop with increasing density in the pore, which are strongly correlated. No sign of thermodynamic phase transition is observed, i.e. the system becomes ordered continuously with increasing density. The close packing structure of the system is a rectangular lattice. In the case of narrower pore three different structures are observed: 1) a fluid with only one layer, 2) a fluid phase with two layers and 3) a solid-like structure with strongly correlating fluid layers. In the wider pore the first structure is missing, while the other two are present. The presence of a solid-like order is due to the strong transversal positional fluctuations and the presence of few defect particles located in the centre of the pore also promotes the formation of solid-like clusters.

The main advantage of our method is that it can be applied for both two- and three-dimensional confined particles with arbitrary particle shape [42]. However the accessible range of the pore width depends strongly on the shape of the confinement and the pair interaction between the particles. The hard confinement can be replaced with periodic boundary condition, too, to mimic the phase behavior of 2D and 3D bulk systems. Percus and Zhang have performed preliminary calculations for hard squares with periodic boundary condition using the transfer matrix method [24], but the comparison with other theories such the fundamental measure theory [43,44] and simulations [45] is still missing. It would be also interesting to compare the crystallization of hard disks with that of hard squares in narrow pores to understand the role of defect particles in the stabilization of the close packing zigzag structure of hard disks and the rectangular structure of hard squares. In hard disk systems the defect particles give rise to jammed states and glassy behavior, while the defects of rectangular structure in the system of hard squares do not slow down the dynamics. To see the structural and dynamical changes more clearly between these two models, it is possible to crossover from square shape to disks one by multi-faceting the squares. In this regard the bulk properties of hard polyhedral particles confined to flat interface has just been examined recently [46].

We believe that the transfer matrix method can be extended for even wider pores with present day computational facilities, too, and shed light on the glassy and crystallization properties of both bulk and confined systems.

## Acknowledgements

We acknowledge the financial support of the Hungarian State and the European Union under the TAMOP-4.2.2.A-11/1/KONV-2012-0071.

**Figures**

**Figure 1.** Schematic representation of four neighboring particles {(*i*,1), (*i*,2), (*i*+1,1) and (*i*+1,2)} are shown, which are represented as hard squares. The neighboring particles are grouped together into two pairs which are labeled as *i* and *i*+1 pairs. The centre-to-centre distance between the neighbors of pairs *i* and *i*+1 are denoted as $x_i$ and $x_{i+1}$, while $X_{i,i+1}$ is the distance between the centers of *i* and *i*+1 pairs. The longitudinal and the transversal directions are denoted as *x* and *y*.

**Figure 2.** Hard squares are confined between two parallel walls. Low-density fluid-like structure is shown in the upper panel, while the solid-like order can be seen in the lower one. The distance between the confining hard walls is $\sigma+W$, where *W* is the effective pore width and $\sigma$ is the side length of the hard squares.

**Figure 3.** Equation of the state of confined hard squares in $P^*$-$\eta$ plane, where $P^* = \beta P \sigma^2 / (\sigma + W)$ is the reduced pressure and $\eta = N\sigma^2 / A$ is the packing fraction. The effective pore width (*W*) is 1.08 $\sigma$ (a) and 1.92 $\sigma$ (b). Exact equation of state (continuous curve) and Tonks equation (short dashed curve) are shown together. Long-dashed vertical line represents the close packing limit.

**Figure 4.** Packing fraction dependence of $<X_1>$ (black curve) and $<X_2>$ (dashed curve) mole fractions at *W*=1.08 $\sigma$ (main panel) and *W*=1.92 $\sigma$ (inset).

**Figure 5.** Transversal positional distribution functions of the hard squares are shown from wall to wall (-$W/2$<$y$<$W/2$) at $W$=1.08 $\sigma$ (upper panel) and $W$=1.92 $\sigma$ (lower panel). Insets show the longitudinal position distribution function in the interval of $0 < x < 2\ \sigma$.

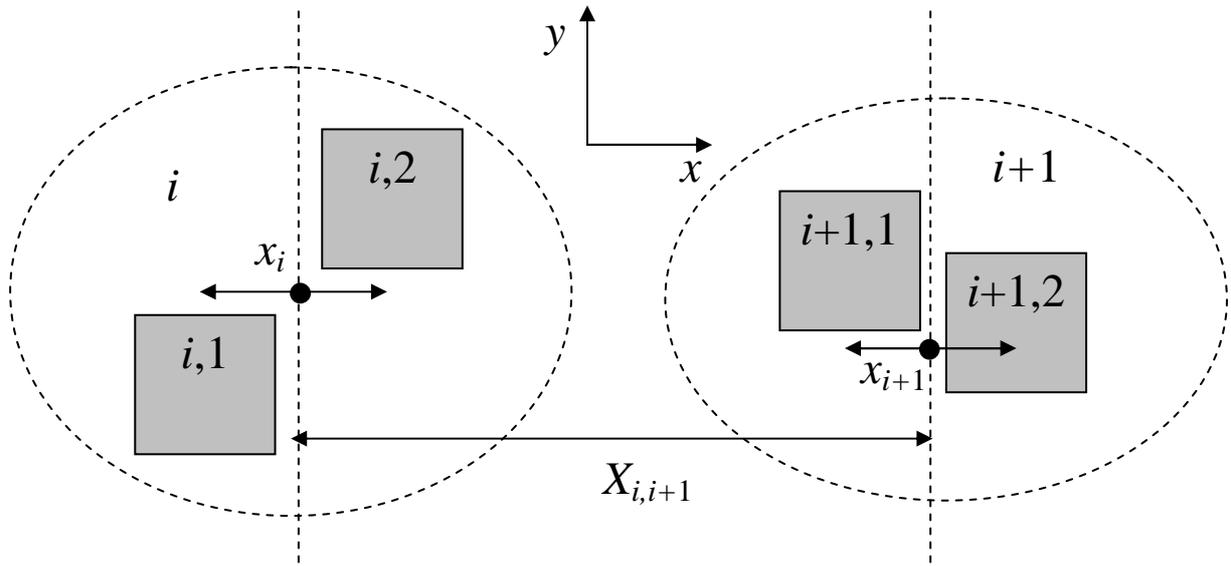

**Figure 1**

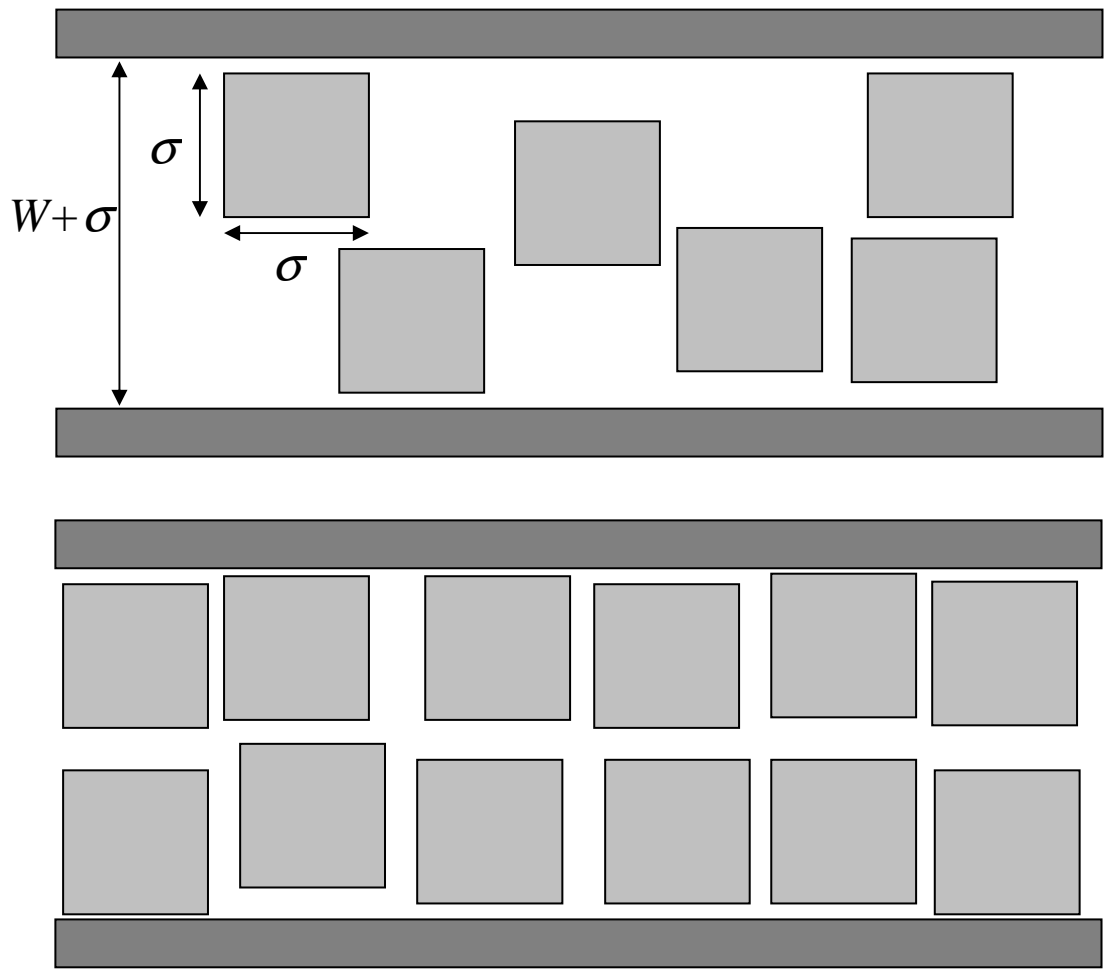

**Figure 2**

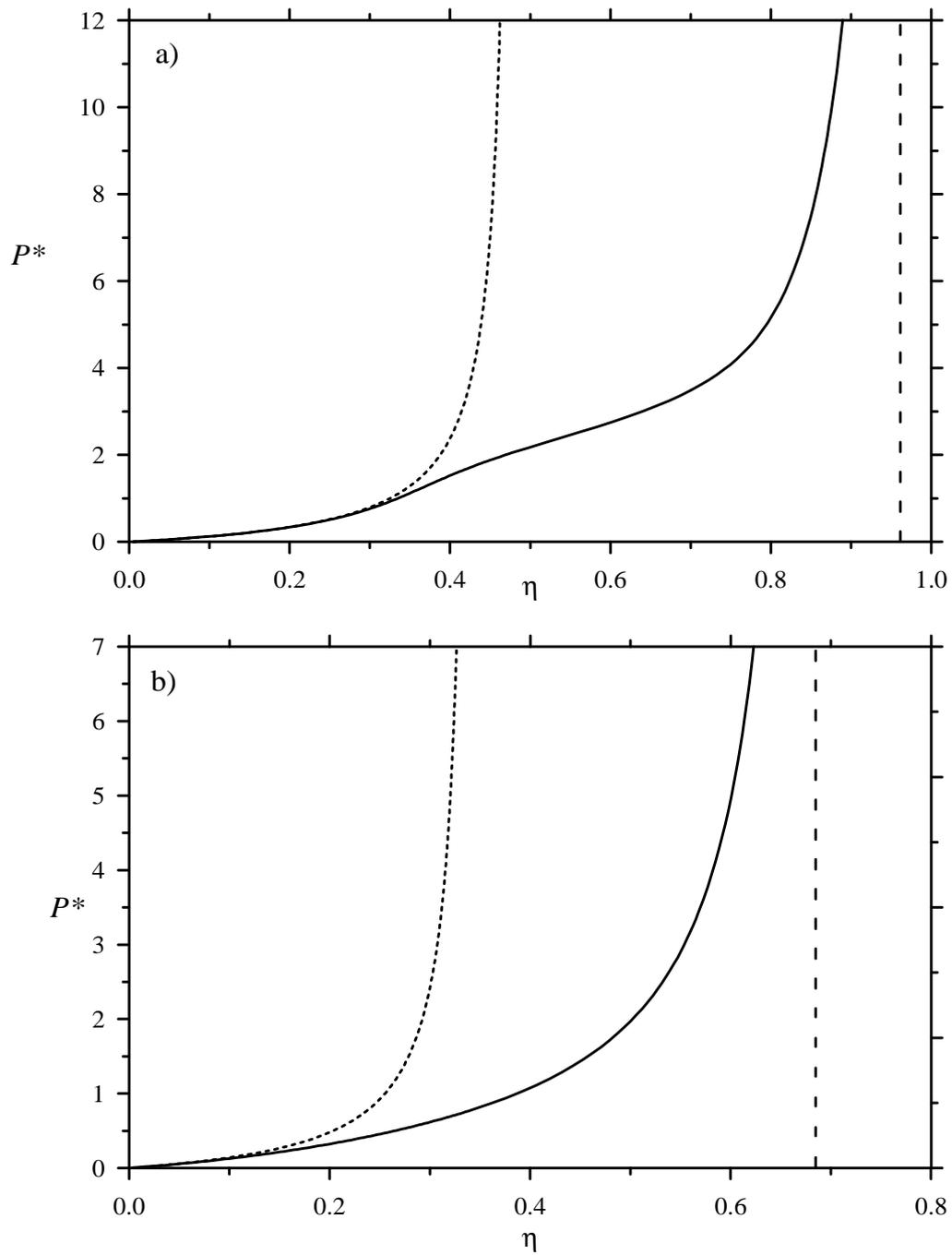

**Figure 3**

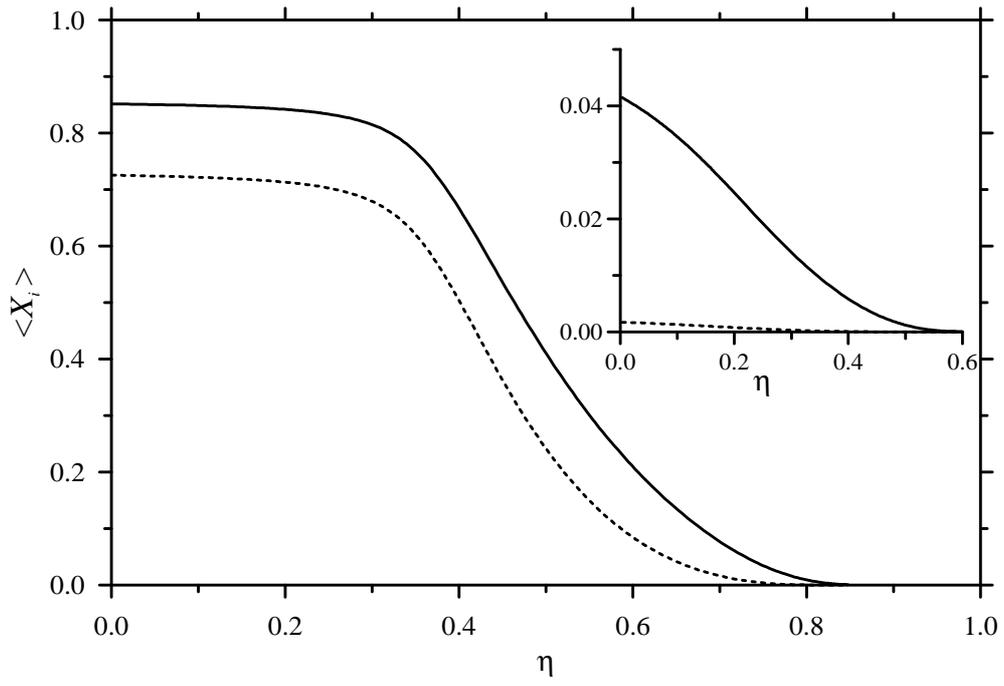

**Figure 4**

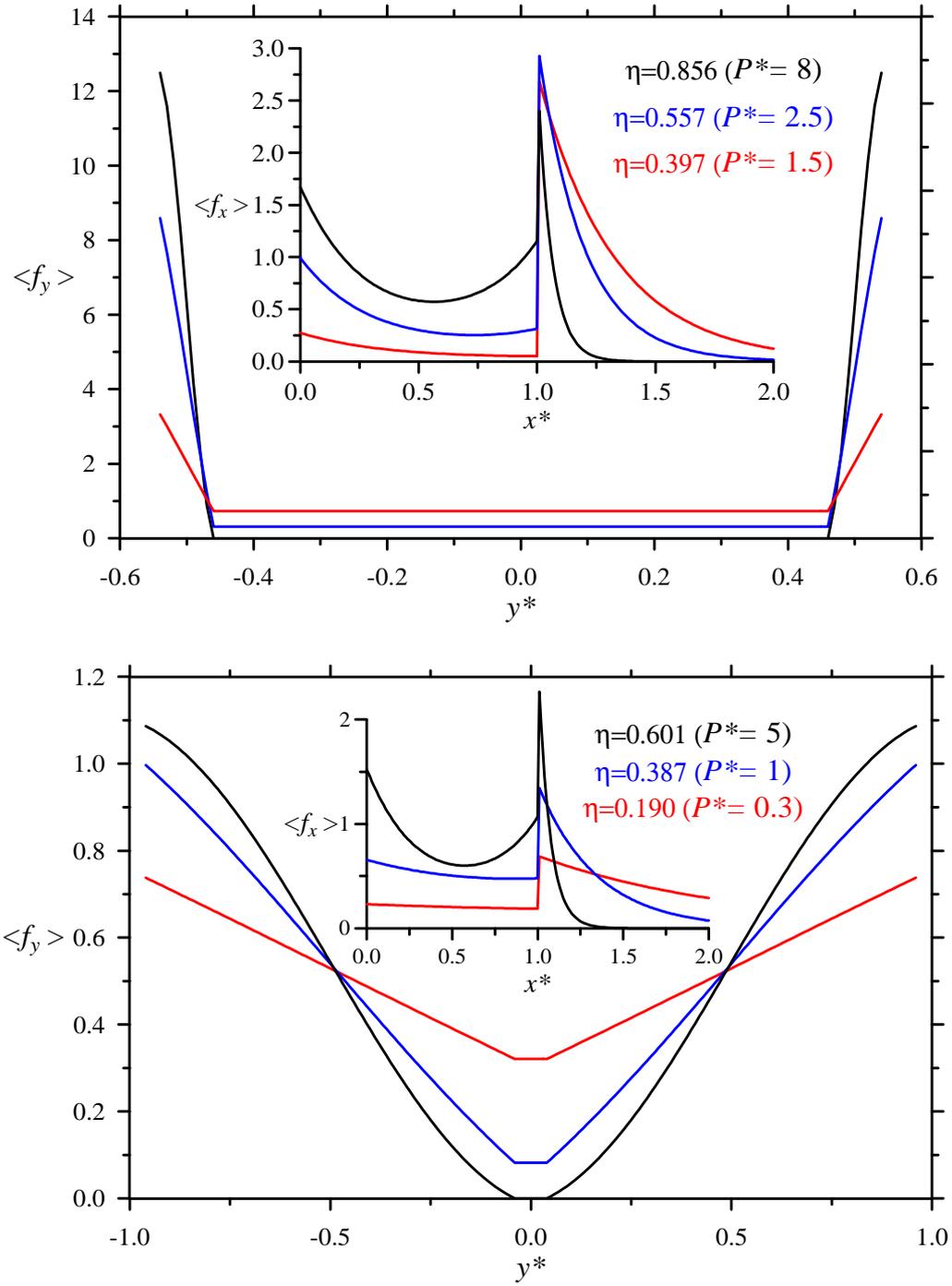

**Figure 5**